\documentclass[conference]{IEEEtran}

\ifCLASSINFOpdf
\else
\fi
\usepackage{amsmath}
\usepackage{amsfonts}
\usepackage[caption=false,font=small,labelfont=sf,textfont=sf]{subfig}
\usepackage[linesnumbered,ruled,vlined]{algorithm2e}
\usepackage{amssymb}
\usepackage{xcolor}
\usepackage{flushend}
\usepackage{comment}
\usepackage{graphicx}
\usepackage{algorithm2e}
\RestyleAlgo{ruled}
\usepackage{algpseudocode}
\usepackage{amsmath}
\usepackage{textcomp}
\usepackage{epsfig}
\usepackage{multicol}
\usepackage{caption}
\usepackage{algpseudocode}
\usepackage{tikz}
\usepackage{geometry}
\usepackage{pgfplotstable}
    \pgfplotsset{
        compat=1.9,
        compat/bar nodes=1.8,
}

\usepackage{acronym}
\acrodef{ce}[CE]{Channel Estimation}
\acrodef{nr}[NR]{New Radio}
\acrodef{graphnet}[GraphNet]{Graph Neural Estimation Network}
\acrodef{mimo}[MIMO]{Multiple Input Multiple Output}
\acrodef{siso}[SISO]{Single Input Single Output}
\acrodef{ls}[LS]{Least Squares}
\acrodef{mmse}[MMSE]{Minimum Mean Square Error}
\acrodef{mse}[MSE]{Mean Square Error}
\acrodef{dl}[DL]{Deep Learning}
\acrodef{cnn}[CNN]{Convolutional Neural Network}
\acrodef{dmrs}[DM-RS]{Demodulation Reference Signals}
\acrodef{ofdm}[OFDM]{Orthogonal Frequency Division Multiplexing}
\acrodef{mmimo}[mMIMO]{Massive MIMO}
\acrodef{gnn}[GNN]{Graph Neural Network}
\acrodef{re}[RE]{Resource Element}
\acrodef{qam}[QAM]{Quadrature Amplitude Modulation}
\acrodef{awgn}[AWGN]{Additive White Gaussian Noise}
\acrodef{prb}[PRB]{Physical Resource Block}
\acrodef{ut}[UT]{User Terminal}
\acrodef{sr}[SR]{Super Resolution}
\acrodef{ir}[IR]{Image Restoration}
\acrodef{srcnn}[SRCNN]{CNN-based SR}
\acrodef{dncnn}[DnCNN]{CNN-based Denoising}
\acrodef{relu}[ReLU]{Rectified Linear Unit}
\acrodef{nn}[NN]{Neural Network}
\acrodef{dnn}[DNN]{Deep Neural Network}
\acrodef{gcn}[GCN]{Graph Convolutional Network}
\acrodef{cdf}[CDF]{Cumulative Distribution Function}
\acrodef{snr}[SNR]{Signal to Noise Ratio}
\acrodef{bs}[BS]{Base Station}
\acrodef{tdl}[TDL]{Tapped Delay Line}
\acrodef{li}[LI]{Linear Interpolation}
\acrodef{mlp}[MLP]{Multi Layer Perceptron}
\acrodef{pdsch}[PDSCH]{Physical Downlink Shared Channel}
\acrodef{tdl}[TDL]{Tapped Delay Line}
\acrodef{bler}[BLER]{Block Error Rate}
\acrodef{cir}[CIR]{Channel Impulse Response}
\acrodef{csi}[CSI]{Channel State Information}
\begin{document}

\pagenumbering{gobble} 
\title{Lightweight Graph Neural Networks for Enhanced 5G NR Channel Estimation}

\author{\IEEEauthorblockN{
{Sajedeh Norouzi\IEEEauthorrefmark{1}$\!\!$, Mostafa Rahmani\IEEEauthorrefmark{2}$\!\!$,  
Yi Chu\IEEEauthorrefmark{2}$\!\!$, Torsten Braun\IEEEauthorrefmark{1}$\!\!$, Kaushik
Chowdhury\IEEEauthorrefmark{3}$\!\!$ , Alister Burr\IEEEauthorrefmark{2}$\!\!$}}
\IEEEauthorblockA{ \IEEEauthorrefmark{1}University of Bern,
\IEEEauthorrefmark{2}University of York,
\IEEEauthorrefmark{3}University of Texas at Austin \\
\{sajedeh.norouzi, torsten.braun\}@unibe.ch;\{m.rahmani,yi.chu,alister.burr\}@york.ac.uk; kaushik@utexas.edu
}
}

%



\maketitle
\begin{abstract}
Effective channel estimation (CE) is critical for optimizing the performance of 5G New Radio (NR) systems, particularly in dynamic environments where traditional methods struggle with complexity and adaptability. This paper introduces GraphNet, a novel, lightweight Graph Neural Network (GNN)-based estimator designed to enhance CE in 5G NR. Our proposed method utilizes a GNN architecture that minimizes computational overhead while capturing essential features necessary for accurate CE. We evaluate GraphNet across various channel conditions, from slow-varying to highly dynamic environments, and compare its performance to ChannelNet, a well-known deep learning-based CE method. GraphNet not only matches ChannelNet’s performance in stable conditions but significantly outperforms it in high-variation scenarios, particularly in terms of Block Error Rate. It also includes built-in noise estimation that enhances robustness in challenging channel conditions.
Furthermore, its significantly lighter computational footprint makes GraphNet highly suitable for real-time deployment, especially on edge devices with limited computational resources. 
By underscoring the potential of GNNs to transform CE processes, GraphNet offers a scalable and robust solution that aligns with the evolving demands of 5G technologies, highlighting its efficiency and performance as a next-generation solution for wireless communication systems.

\end{abstract}


\IEEEpeerreviewmaketitle
\section{Introduction}
The deployment of 5G \ac{nr} systems represents a major advancement in telecommunications, delivering high-speed, low-latency, and reliable connectivity crucial for big data throughput and real-time responsiveness, while also enabling more sustainable digital infrastructure through enhanced energy efficiency and optimized network resource utilization \cite{ahmadi2025towards}. Central to optimizing these advanced networks is the process of \ac{ce}, which predicts \ac{csi} from received pilot signals. This process is vital as channel characteristics can distort the received signal, requiring accurate estimation and compensation to recover the transmitted symbol effectively.
\ac{ce} in 5G \ac{nr} typically begins with pilot symbols known to both the transmitter and receiver, positioned within time-frequency grids. Traditional methods like \ac{ls} and \ac{mmse} balance performance against computational complexity. However, these methods often struggle in scenarios characterized by low \acp{snr} or rapidly changing channel conditions. Furthermore, the sparse pilot signal arrangement, particularly under high Doppler spreads, presents substantial challenges for accurate \ac{ce} in mobile contexts due to the inadequacy of traditional interpolation techniques in these dynamic environments.
Recent advancements have seen the adoption of \ac{dl} techniques to improve \ac{ce}, addressing the limitations inherent in traditional methods, especially in handling sparse pilot distributions and fluctuating channel conditions. Notably, 2D \ac{cnn}-based channel refinement techniques \cite{Soltani} and algorithms utilizing recurrent and residual neural networks \cite{9217192, Li} have been explored. Conditional generative adversarial networks have also been applied to \ac{ce}, leveraging received signals as conditional inputs \cite{ye2022channel}. Despite their potential, these \ac{dl} approaches often incur high computational overheads, challenging their real-time deployment feasibility.

Recent methodologies have been proposed to optimize \ac{dl} models to mitigate these computational demands, reducing their complexity and energy consumption while still aiming to balance high accuracy with reduced inference time \cite{10770678}. However, these improvements predominantly rely on \ac{sr} techniques, leaving the potential of \ac{gnn}-based CE underexplored. \acp{gnn}, known for their efficacy in modeling complex relationships within wireless communication systems \cite{shen2022graph}, offer promising new avenues for research. They have shown notable potential in enhancing CE generalization across varying training and testing scenarios by learning mappings between estimated and true channel matrices, maintaining robust performance regardless of antenna count due to their permutation equivariance \cite{Mingye}.
Furthermore, GNNs have been applied to capture intricate relationships between antennas in multi-user systems \cite{10460648}, and in combination with CNN layers, to manage multiple users within 5G NR MIMO systems, effectively reducing computational complexity \cite{cammerer2023neural}. A GNN-based approach has also been proposed to track channel information in mMIMO networks with high mobility \cite{yang2020graph}, demonstrating the extensive applicability of GNNs beyond traditional DL methods. Unlike SR or CNN techniques that enhance initial estimations using adjacency in a grid, GNNs can aggregate more accurate information directly from DM-RSs, thereby significantly enhancing CE accuracy. This paper will address the following key research questions to thoroughly evaluate the application of GNNs in 5G \ac{ce}:
\begin{enumerate}
    \item What are the optimal design strategies for a GNN architecture tailored for CE, and how does it perform relative to existing methods?
    \item How can concurrent noise estimation be integrated within CE processes?
    \item How does the proposed algorithm perform in practical scenarios? 
\end{enumerate}
To address these research questions, we introduce and evaluate \ac{graphnet}: The main contributions are as follows: 
\begin{enumerate}
    \item 
    By defining nodes, edges, and features of the wireless channel graph, we enable the application of \acp{gnn} to \ac{ce}. A \ac{gnn} architecture is designed to estimate channels and its performance is compared to established methods, such as \ac{ls}, practical \ac{ce}, and a \ac{dl}-based algorithm \cite{Soltani}. 
    \item A noise estimation module is incorporated directly within the hidden layers of \ac{graphnet}, leveraging the latent node embeddings. This dual estimation approach improves \ac{bler} performance without additional overhead in model complexity. 
    \item 
    We validate our methodology using a MATLAB-based 5G simulation framework, demonstrating its theoretical and practical robustness. The data is generated using a simulation setup incorporating realistic channel conditions following the 3GPP NR specifications. Compared to other algorithms, the proposed method performs competitively and, in some cases, better with significantly lighter model size.
\end{enumerate}
By representing the wireless channel as a graph and integrating \ac{gnn}-based \ac{ce} with simultaneous noise estimation, the proposed \ac{graphnet} framework enhances CE performance while reducing computational complexity. Simulation results confirm its practical applicability in realistic scenarios, demonstrating competitive performance compared to other DL-based approaches, often with significantly lower model complexity. Furthermore, \ac{graphnet} has been explored in alternative configurations, including joint \ac{ce} and interpolation, as well as a federated learning-based approach across different \acp{bs}. These extensions highlight its adaptability and potential for future wireless communication systems \cite{norouzi2025decentralized}.

\section{system model}
We consider the downlink \ac{siso} \ac{ofdm} system model for 5G-\ac{nr}. In an OFDM system, $M$ represents the number of subcarriers, and $N$ represents the number of symbols forming the resource grid. Let $\mathbf{x} \in \mathbb{C}^{M \times N}$ represent a block of the complex-valued transmitted baseband signal. In block $\mathbf{x}$, some \acp{re} are allocated for known pilots instead of data, known as \ac{dmrs}. We denote the pilot positions by $p \in \mathcal{P}$, where $p = (p_m, p_n)$ defines the indices of pilot positions in the resource grid. In this case, we assume \ac{ofdm} with a sufficiently long cyclic prefix. Therefore, $\mathbf{y}_{m, n}$ represents the received signal for the \ac{re} $(m,n)$ and can be defined as follows.
\vspace{-0.125 cm }
\begin{equation}
    \label{yreceived}
    \mathbf{y}_{m, n} = \mathbf{h}_{m, n} \mathbf{x}_{m, n} + \mathbf{w}_{m, n},
\end{equation}
where $\mathbf{h}_{m, n}$ is the wireless channel, and $\mathbf{w}_{m, n} \sim \mathcal{CN}(0, \sigma^2)$ is the complex-valued \ac{awgn} with noise power $\sigma^2$. 
\begin{figure}
    \centering
    \includegraphics[width=0.8\linewidth]{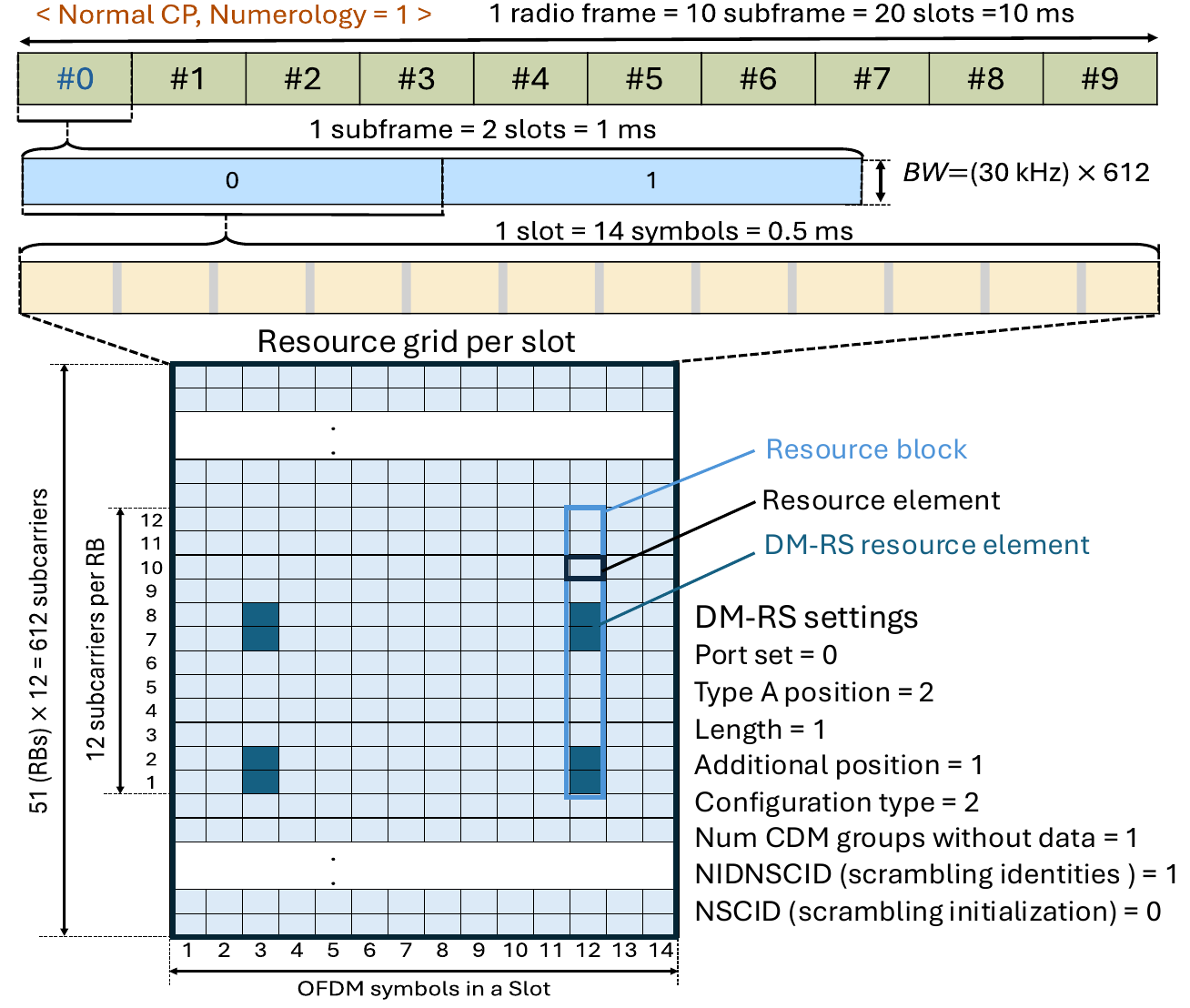}
    \caption{5G NR frame structure including carrier, PDSCH, and DM-RS configuration}
    \label{frame}
\end{figure}
\subsection{Resource Grid Configuration}
In this system, each frame comprises 10 subframes, each of duration 1\,\text{ms} and partitioned into 20 slots, based on 5G \ac{nr} numerology $\mu =1$. The subcarriers are grouped into \acp{prb}, with each \ac{prb} consisting of 12 consecutive subcarriers. For this configuration, we allocate 51 \acp{prb}, corresponding to 612 subcarriers. The system operates per slot, comprising 14 consecutive \ac{ofdm} symbols. Certain OFDM symbols are reserved for pilot signals, essential for \ac{ce}. \ac{dmrs} symbols occur at fixed positions within each slot, specifically at the 3rd and 12th \ac{ofdm} symbols. 

While the positions of the \ac{dmrs} remain static across slots, the pilot symbol values depend on the specific slot number within a frame, allowing for efficient \ac{ce} over time-varying channels. Each slot thus forms a resource grid of 612 subcarriers over 14 OFDM symbols, where 8 out of 12 subcarriers in each grid, four at symbol 3 and 4 at symbol 12, are reserved for \ac{dmrs}, and the remaining are available for data transmission (Fig. \ref{frame}). This resource grid represents the transmission of an entire transport block, which the receiver decodes based on the slot's received signal. The transmission operates per slot, processing the received resource grid to recover the transport block. Below, we will explore traditional and practical CE methods used in 5G-NR alongside state-of-the-art \ac{dl} techniques. We will also introduce our proposed method, which serves as the primary focus of this research.

\subsection{Least Square Method \& Linear Interpolation} \label {LSLI}
In this sub-section, we present \ac{ls} with \ac{li} as a conventional estimation method and use it as a baseline for comparison. For every \ac{re} carrying a pilot, we can obtain a complex-valued channel estimate $\hat{\mathbf{h}}_{m_p, n_p}$ according to 
\begin{align}
    \hat{\mathbf{h}}_{m_p, n_p}= \frac{\mathbf{p}_{m_p, n_p}\hat{\mathbf{y}}_{m_p, n_p}}{|\mathbf{p}_{m_p, n_p}|^2},
\end{align}
where $\mathbf{p}_{m_p, n_p} \in \mathbb{C}$ is the pilot signal transmitted by the transmitter on the \ac{re} $(m_p, n_p)$ and $\hat{\mathbf{y}}_{m_p, n_p}$ is the received pilot signals. After estimating values for known pilot signals, the 2D \ac{li} method can be applied to estimate the channel transfer function for non-\ac{dmrs} \acp{re} based on the values from adjacent \ac{dmrs} \acp{re}. This approach ensures that closer \ac{dmrs} \acp{re} contribute more to the estimated value, ensuring they profoundly influence the estimated transfer function. 

\subsection{Practical Channel Estimation}
The practical CE method, commonly known as the 5G-NR practical estimator in MATLAB, begins with LS estimation using received \ac{dmrs} symbols, providing an initial but noise-sensitive channel estimate. \ac{cir}-based denoising enhances accuracy, removing insignificant noise components while preserving dominant channel taps. Noise averaging further improves estimation by reducing random variations over time and frequency. Finally, LI in both the frequency and time domains reconstructs the complete channel response, ensuring a smooth and accurate estimation. While the 5G-NR practical estimator offers a balance between accuracy and efficiency, it suffers from several limitations: sensitivity to noise, performance degradation under high mobility, dependency on reference signal quality and density, and increased computational load in certain configurations \cite{ghourtani2024link}. 

\subsection{ChannelNet} 
ChannelNet utilizes image processing techniques such as \ac{sr} and \ac{ir}, which focus on reconstructing high-resolution images from low-resolution inputs and image denoising \cite{Soltani}. The approach begins by estimating the channel at pilot positions, then interpolating and treating it as a low-resolution image. It comprises two cascaded networks; the first stage employs \ac{srcnn}, which enhances \ac{ce} by modeling the channel as an image, with its real and imaginary components represented as two distinct images. The second stage uses the \ac{dncnn} for further refinement. The ultimate goal is to reconstruct a high-resolution image, yielding a more accurate channel estimate. Mathematically, this \ac{ce} problem can be expressed as follows:
\begin{equation}
    \bar{\mathbf{h}}_{m,n} = f_{\text{IR}}(f_{\text{SR}} (\hat{\mathbf{h}}_{m_p,n_p} ; \Theta_\text{SR}) ; \Theta_{\text{IR}}),
\end{equation}
where $f_{\text{SR}}$ and $f_{\text{IR}}$ represent the \ac{sr} and \ac{ir} functions, which are parameterized by $\Theta_\text{SR}$ and $\Theta_{\text{IR}}$, respectively. The network is optimized by minimizing the \ac{mse} between the estimated channel by ChannelNet ($\bar{\mathbf{H}}$) and the actual channel as follows. 
\begin{equation}
    \mathcal{L} (\Theta_{\text{SR}}, \Theta_{\text{IR}})  = \frac{1}{T} \sum_{t \in T} \| \bar{\mathbf{H}}(t) - \mathbf{H} (t)\|_2^2, 
\end{equation}
where $\|.\|_2$ is the Euclidean norm and $T$ is the number of training samples.
\section{GraphNet for Channel and Noise Estimation}
In this subsection we present \ac{graphnet}. The approach models the channel as a graph where each \ac{re} is treated as a node, and edges are formed based on the proximity of these nodes to \ac{dmrs} nodes. This graph structure is leveraged to enhance channel transfer function estimation and noise estimation through a GraphSAGE, an inductive \ac{gnn} framework \cite{hamilton2017inductive}. Unlike traditional transductive \ac{gnn} methods that require the entire graph during training, GraphSAGE learns a generalizable aggregation function that generates node embeddings for previously unseen nodes by sampling and aggregating features from their local neighborhoods. 
\subsection{Graph Representation of the Wireless Channel}
\label{graphrep}
Each node in the graph corresponds to a \ac{re}, which is defined as a unique subcarrier-symbol pair. Thus, each node $v_i \in \mathcal{V}$ represents a unique pair $(m_i, n_i)$, where $m_i$ is the subcarrier index and $n_i$ is the symbol index. The set of nodes is defined as: $\mathcal{V} = \{ v_1, v_2, \dots, v_V \}$, where $V = M \times N$ is the total number of nodes in the graph, corresponding to the total number of \acp{re}. The edge set $\mathcal{E}$ defines the connections between the nodes in the graph. Each node $v_i$ is connected to its four closest \ac{dmrs} nodes based on the Euclidean distance in the frequency-time grid. For each non-\ac{dmrs} node $v_i$, we select the four closest \ac{dmrs} nodes $\{ v_j^{p} \mid j \in \mathcal{N}_{p}(v_i) \}$ where $\mathcal{N}_{p}(v_i)$ represents the set of the four nearest ones. Hence, the edge set $\mathcal{E}$ is defined as: $\mathcal{E} = \{ (v_i, v_j^{p}) \mid i \neq j, v_i \in \mathcal{V}, v_j^{p} \in \mathcal{N}_{p}(v_i) \}$ (Alg. \ref{graphconstruction}).

\begin{algorithm}[hbt!] 
\footnotesize
\caption{Graph Construction from OFDM Grid}\label{graphconstruction}
\KwData{$M$ subcarriers, $N$ symbols, \ac{dmrs} positions $\mathcal{P}$}
\KwResult{Node mappings $\mathcal{V}$, Reverse mapping $\mathcal{V}^{-1}$, Edge index $\mathcal{E}$} 
\tcc{\scriptsize  Create node mapping}
$\mathcal{V} \gets \{\}, \mathcal{V}^{-1} \gets \{\}$\;
$\text{node\_idx} \gets 0$\;
\For{$m = 0$ \KwTo $M-1$}{
    \For{$n = 0$ \KwTo $N-1$}{
        $\mathcal{V}[i] \gets (m, n)$ \tcc{ \scriptsize Node to grid mapping}
        $\mathcal{V}^{-1}[(m, n)] \gets i$ \tcc{\scriptsize Reverse mapping}
        $i \gets i + 1$\;
    }
}
\tcc{\scriptsize Convert \ac{dmrs} positions to node indices}
$\mathcal{N}_{p} \gets []$\;
\ForEach{$ p = (m_p, n_p) \in \mathcal{P}
$}{
    $\mathcal{N}_{p} \leftarrow \mathcal{N}_{p} \cup  \mathcal{V}^{-1}[(m_p, n_p)] $\;
}
\tcc{\scriptsize Create edges based on proximity to \ac{dmrs}}
$\mathcal{E} \gets []$\;
\For{$v_i = 0$ \KwTo $|\mathcal{V}|-1$}{
    $(m_i, n_i) \gets \mathcal{V}[i]$ \tcc{\scriptsize Get grid position of node i}
    $\mathbf{d} \gets []$ \tcc{\scriptsize Distance array}
    
    \ForEach{$v_j \in \mathcal{N}_{p}$}{
        $(m_j, n_j) \gets \mathcal{V}[j]$\;
        $d_{ij} \gets \sqrt{(m_i - m_j)^2 + (n_i - n_j)^2}$ \tcc{\scriptsize Euclidean distance}
        $\mathbf{d} \cup (d_{ij}, j)$\;
    }
    
    $\mathbf{d}^{\uparrow} \gets \text{sort}(\mathbf{d})$ \tcc{\scriptsize Sort by distance}
    $\mathcal{N}_p(v_i) \gets \mathbf{d}_{sorted}[0:3]$ \tcc{\scriptsize Select 4 nearest DM-RS}
    
    \ForEach{$(d, j) \in \mathcal{N}_p(v_i)$}{
        $\mathcal{E} \cup (v_i, v_j)$ \tcc{\scriptsize Add edge from node $v_i$ to \ac{dmrs} node $v_j$}
    }
}

\Return{$\mathcal{V}, \mathcal{V}^{-1}, \mathcal{E}$}
\end{algorithm}

\subsection{Channel Estimation via GraphSAGE Aggregation}
The node feature aggregation in \textit{GraphSAGE} follows a neighborhood-based aggregation scheme. The first layer aggregation is computed as:
\begin{align} \label{outfirstlayer}
    \tilde{\mathbf{h}}_i^{(1)} = \sigma ( \mathbf{W}_1 \mathbf{\phi}_i + \sum_{j \in \mathcal{N}(v_i)} \mathbf{W}_2 \mathbf{\phi}_j ),
\end{align}
where $\mathbf{\phi}_i \in \mathbb{R}^{F_{\text{in}}}$ is the feature vector of node $v_i$ and is initialized from $\hat{\mathbf{h}}_{m,n}$ (subsection \ref{LSLI}), $F_{\text{in}}$ equals to two, which represents the real and imaginary component of $\hat{\mathbf{h}}_{m,n}$. $\mathcal{N}(v_i)$ denotes the set of neighboring nodes of $v_i$, $\mathbf{W}_1, \mathbf{W}_2 \in \mathbb{R}^{F\times F_{\text{in}}}$ are learnable weight matrices, where  $F$ is dimensionality of hidden layer node features and $\sigma(.)$ is the \ac{relu} activation function ensures non-linearity in feature transformations. This process is repeated for additional aggregation layers. The hidden-layer aggregation is defined as:

\begin{align} \label{outsecondlayer}
    \tilde{\mathbf{h}}_i^{(2)} = \sigma ( \mathbf{W}_3 \tilde{\mathbf{h}}_i^{(1)} + \sum_{j \in \mathcal{N}(v_i)} \mathbf{W}_4 \tilde{\mathbf{h}}_j^{(1)} ),
\end{align}
where $\mathbf{W}_3, \mathbf{W}_4 \in \mathbb{R}^{F \times F}$ are learnable weight matrices associated with the hidden layer. The final estimated channel transfer function for node $v_i$ is computed as $\tilde{\mathbf{h}}_i = \mathbf{W}_5 \tilde{\mathbf{h}}_i^{(2)}$, where $\mathbf{W}_5 \in \mathbb{R}^{F \times F\text{out}}$ is the learnable weight matrix of the output layer and $F_{\text{out}} =2$ is the expected feature dimension for each node. The two output features represent the real and imaginary parts of the channel, respectively. Specifically, if we denote the output vector as $\tilde{\mathbf{h}}_i = [\tilde{\mathbf{h}}_i[0], \tilde{\mathbf{h}}_i[1]]$, then the estimated complex channel for node $v_i$ is given by 
\begin{align}
    \tilde{\mathbf{h}}_i = \tilde{\mathbf{h}}_i[0] + j.\tilde{\mathbf{h}}_i[1],
\end{align}
where $\tilde{\mathbf{h}}_i[0]$ and $\tilde{\mathbf{h}}_i[1]$ correspond to real and imaginary parts, respectively. By aggregating the estimated channels for all nodes, we obtain the full channel estimate. 

\subsection{Noise Estimation via Global Embedding Aggregation}
\begin{figure}
    \centering
    \includegraphics[width=0.8\linewidth]{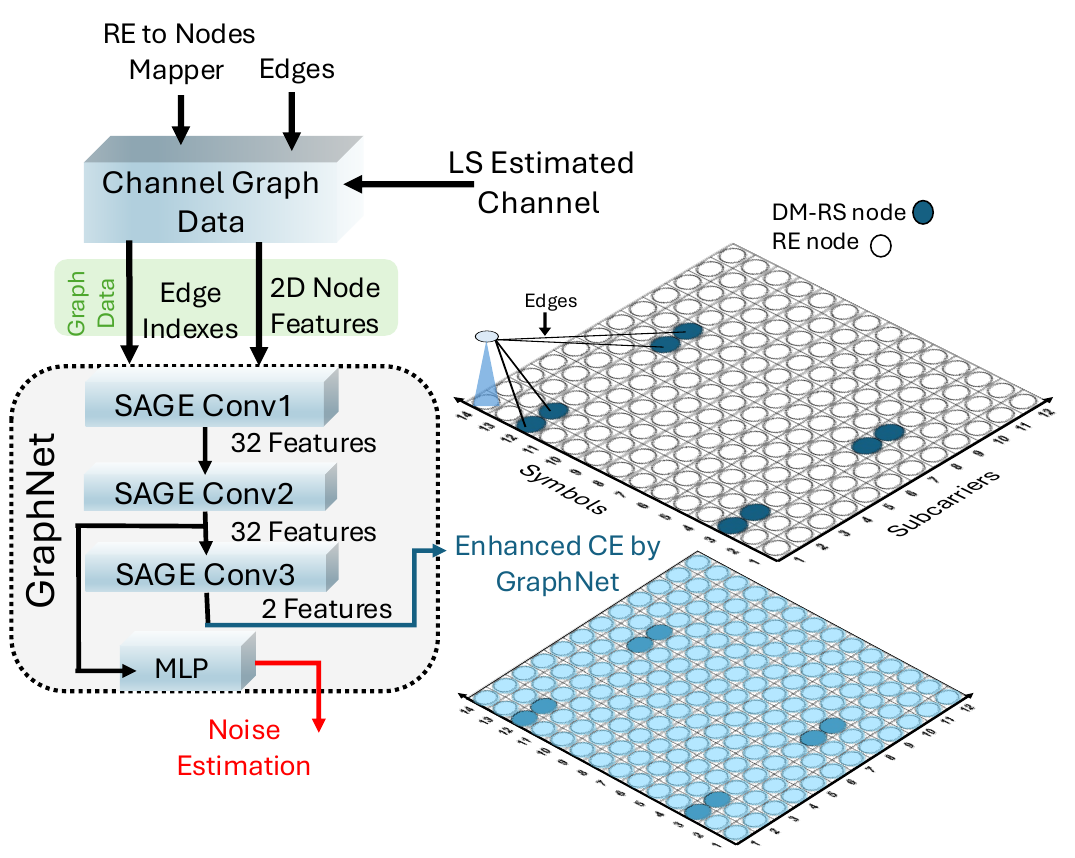}
    \caption{GraphNet-based CE algorithms}
    \label{Fig_1}
\end{figure}

In addition to estimating $h$, the model predicts the noise affecting the channel. Instead of computing noise at the individual node level, the model aggregates hidden node representations across the entire graph to obtain a noise estimation. First, the hidden embeddings from the second GraphSAGE layer are aggregated across all nodes:
\begin{align} \label{noiseeq}
    \mathbf{h}_{\gamma} = \frac{1}{V} \sum_{i=1}^{V} \tilde{\mathbf{h}}_i^{(2)},
\end{align}
where $\tilde{\mathbf{h}}_i^{(2)} \in \mathbb{R}^{F}$ is the output of the second GraphSAGE layer of node $i$, and $V$ is the number of nodes in the graph. To estimate the noise, this global representation $\mathbf{h}_{\gamma}$ is passed through a two-layer \ac{mlp}: 
$\hat{n} = \mathbf{W}_7 (\sigma (\mathbf{W}_6 \mathbf{h}_{\gamma} + \mathbf{b}_6)) + \mathbf{b}_7 $ where, $\mathbf{W}_6 \in \mathbb{R}^{F \times F_m}, \mathbf{b}_6 \in \mathbb{R}^{d_m}$ (hidden layer transformation), $\mathbf{W}_7 \in \mathbb{R}^{1 \times d_m}, \mathbf{b}_7 \in \mathbb{R}$ (final noise estimation). For each channel sample \ac{graphnet} jointly outputs the estimated channel function for each node and noise estimation for the entire graph; the final model output is:
\begin{align}
    \text{\ac{graphnet}}(\mathbf{\Phi}, \mathcal{E}) = \left( \tilde{\mathbf{H}}, \hat{n} \right),
\end{align}
where $\tilde{\mathbf{H}} = [\tilde{\mathbf{h}}_1, \tilde{\mathbf{h}}_2, ..., \tilde{\mathbf{h}}_V]$ represents the estimated channel values for all nodes and $\hat{n}$ is the noise estimate. The network is optimized by minimizing the \ac{mse} loss consisting of two components; \ac{ce} loss $\mathcal{L}_{\text{CE}}$ and noise estimation loss $\mathcal{L}_{\text{no}}$:
\begin{align} 
    \mathcal{L}_{\text{total}} = \lambda_{\text{CE}} \frac{1}{V} \sum_{i=1}^{V} \| \tilde{\mathbf{h}}_i - \mathbf{h}_i \|^2_2 + \lambda_{\text{no}}  \left( \hat{n} - n \right)^2. 
\end{align}
$\lambda_{\text{CE}}$ and $\lambda_{\text{no}}$ are weighting factors and can be adjusted to prioritize either \ac{ce} or noise estimation depending on the application requirements. This approach ensures that the model effectively captures both local node dependencies for \ac{ce} and global graph properties for noise estimation, leading to improved performance in wireless communication tasks. The proposed GNN-based CE process is illustrated in Fig. \ref{Fig_1} and Alg. \ref{alg:graphsage_forward}, \ref{alg:graphsage_train_test}. In the first step, graph data is constructed as described in Subsection \ref{graphrep}. This data is then processed by the GraphNet architecture, which comprises three SAGE convolution layers for CE and a \ac{mlp} layer dedicated to noise estimation. 

\begin{algorithm}[hbt!]
\small
\caption{GraphNet Model Architecture}\label{alg:graphsage_forward}
\KwData{Graph data $\mathcal{G} = (\mathbf{\phi}, \mathcal{E})$, $\mathbf{\phi} \in \mathbb{R}^{|\mathcal{V}| \times 2}$ (node features), $\mathcal{E} \in \mathbb{Z}^{2 \times |\mathcal{E}|}$ (edge index)}
\KwResult{Channel output $\tilde{\mathbf{H}} \in \mathbb{R}^{|\mathcal{V}| \times 2}$, Noise estimate $\hat{n} \in \mathbb{R}$}

\tcc{\scriptsize Three-layer GraphSAGE convolution}
$\tilde{\mathbf{H}}^{(1)} \gets \sigma(\text{SAGEConv}_1(\mathbf{\phi}, \mathcal{E}))$ (eq. \ref{outfirstlayer}) \tcc{\scriptsize First SAGE layer: $2 \rightarrow 32$} 
$\tilde{\mathbf{H}}^{(2)} \gets \sigma(\text{SAGEConv}_2(\tilde{\mathbf{H}}^{(1)}, \mathcal{E}))$ (eq. \ref{outsecondlayer})\tcc{\scriptsize Second SAGE layer: $32 \rightarrow 32$}
$\tilde{\mathbf{H}} \gets \text{SAGEConv}_3(\tilde{\mathbf{H}}^{(2)}, \mathcal{E})$ \tcc{\scriptsize Output layer: $32 \rightarrow 2$}

\tcc{\scriptsize Global noise estimation} 
${\mathbf{H}}_\gamma \gets \frac{1}{|\mathcal{V}|} \sum_{v=1}^{|\mathcal{V}|} \tilde{\mathbf{H}}^{(2)}$ (eq. \ref{noiseeq}) \tcc{\scriptsize Mean pooling of hidden features}
$\hat{n} \gets \text{MLP}_{noise}(\mathbf{H}_\gamma)$ \tcc{\scriptsize MLP: $32 \rightarrow 32 \rightarrow 1$}

\Return{$\tilde{\mathbf{H}}, \hat{n}$}
\end{algorithm}

\begin{algorithm}[hbt!]
\caption{GraphNet Training \& Evaluation}\label{alg:graphsage_train_test}
\footnotesize
\KwData{Train data $\mathbf{D} \in \mathbb{R}^{T \times M \times N \times 2}$, Test data $\mathbf{D}_t$, Channel labels $\mathbf{H}, \mathbf{H}_t$, Noise labels $n, n_t$, \ac{dmrs} positions $\mathcal{P}$, Epochs $E$}
\KwResult{Trained model $\mathcal{M}$, predictions $\tilde{\mathbf{H}}, \hat{n}$, test losses}

\tcc{\scriptsize Initialize model and optimizer}
$\mathcal{M} \gets \text{GraphSAGE}(2, 32, 2)$; \quad $\text{optimizer} \gets \text{Adam}(\mathcal{M}, 0.001)$;

\tcc{\scriptsize Node mapping and edge setup}
$\mathcal{V}, \mathcal{V}^{-1} \gets \text{CreateNodeMapping}(M, N)$; \quad $\mathcal{E} \gets \text{PrecomputeEdges}(\mathcal{V}, \mathcal{V}^{-1}, \mathcal{P})$;

\tcc{\scriptsize Graph conversion}
$\mathcal{G} \gets \text{CreateGraphData}(|\mathcal{V}|, \mathcal{V}, \mathcal{E}, \mathbf{D})$; \quad $\mathcal{G}_t \gets \text{CreateGraphData}(|\mathcal{V}|, \mathcal{V}, \mathcal{E}, \mathbf{D}_t)$;

\tcc{\scriptsize Training}
\For{$e = 1$ \KwTo $E$}{
    \For{$t = 1$ \KwTo $T$}{
        $\tilde{\mathbf{H}}[t], \hat{n}[t] \gets \mathcal{M}(\mathcal{G}[t])$;

        $\mathcal{L}_{\text{CE}} \gets \text{MSE}(\tilde{\mathbf{H}}[t], \mathbf{H}[t])$; \quad $\mathcal{L}_{\text{no}} \gets \text{MSE}(\hat{n}[t], n[t])$;

        $\mathcal{L} \gets \lambda_{\text{CE}}\mathcal{L}_{\text{CE}} + \lambda_{\text{no}}\mathcal{L}_{\text{no}}$;

    }
    $\text{Save Trained Model }\mathcal{M}$;
}

\tcc{\scriptsize Evaluation}
$\mathcal{M}.\text{eval()}$; \quad $\mathcal{L}_{\text{CE}}, \mathcal{L}_{\text{no}}, \mathcal{L}_{\text{total}} \gets 0$;

\For{$t = 1$ \KwTo $T_{test}$}{
    $\tilde{\mathbf{H}}[t], \hat{n}[t] \gets \mathcal{M}(\mathcal{G}_t[t])$;

    $\mathcal{L}_{\text{CE}} \gets \text{MSE}(\tilde{\mathbf{H}}[t], \mathbf{H}_t[t])$; \quad $\mathcal{L}_{\text{no}} \gets \text{MSE}(\hat{n}[t], n_t[t])$;
}

\Return{$\mathcal{M}, \tilde{\mathbf{H}}, \hat{n}, \bar{\mathcal{L}}_{\text{CE}}, \bar{\mathcal{L}}_{\text{no}}, \bar{\mathcal{L}}_{\text{total}}$}
\end{algorithm}

\section{Performance Evaluation}
\subsection{Data Generation}

\textcolor{black}{The training data for this study is generated using a simulation framework built on the 5G MATLAB Toolbox. Key channel parameters, including the \ac{snr}, Doppler shift, and delay spread, are systematically varied to produce a diverse and representative dataset of realistic wireless propagation environments. The channel model complies with the 3GPP \ac{nr} specifications, specifically utilizing the \ac{tdl} model, which accurately captures multi-path and time-varying characteristics typical of real-world channels. The simulation framework incorporates the configuration of \ac{nr} carrier parameters such as bandwidth, subcarrier spacing, and cyclic prefix, alongside antenna configurations and \ac{pdsch} settings. Multiple \ac{tdl} profiles (TDL-A to TDL-E) are employed to represent varying propagation conditions. For each generated sample, the delay spread is randomly selected from a predefined range in nanoseconds, while the Doppler shift is drawn from a specified frequency range in Hertz, ensuring a wide coverage of channel dynamics.}

\textcolor{black}{The simulated waveform adopts a numerology with a subcarrier spacing of 30kHz (corresponding to $\mu = 1$), and normal cyclic prefix, consistent with typical Frequency Range 1 (FR1) deployments in 5G \ac{nr} systems. The \ac{pdsch} is allocated over the entire available bandwidth and spans the full duration of the slot, enabling the utilization of the complete time-frequency grid. The \ac{dmrs} configuration employs Type 2, where \ac{dmrs} symbols are inserted starting at the third \ac{ofdm} symbol of each slot, with an additional position enabled to enhance channel tracking capabilities, particularly under conditions with significant Doppler spread or multi-path delay.}

Using these configurations, a transmitted waveform is generated and passed through a randomly parameterized TDL channel. Additive white Gaussian noise is introduced based on a random \ac{snr} value, with noise power calculated according to \ac{snr}, the number of receive antennas, and carrier-specific parameters. \ac{dmrs} indices and symbols are computed and used to populate the resource grid, which is modulated via \ac{ofdm}. From the received signal, \ac{dmrs} symbols are extracted, and a \ac{li} is applied to estimate the full channel response across the grid in both real and imaginary components, accounting for the complex-valued nature of the channel. Perfect channel estimates—computed using path gains and filters from the simulation—serve as ground truth labels. Each data sample consists of the estimated channel grid as input and the corresponding perfect channel estimate as the label. This process is repeated over numerous channel realizations with randomized parameters to ensure a diverse and robust training dataset.

\begin{table}
    \centering
    \footnotesize
    \caption{PARAMETERS FOR DATASET}
    \begin{tabular}{|l|l|} \hline
        \textbf{Parameter} & \textbf{Value/Description}\\ \hline
        Channel Profiles & TDL-A, TDL-B, TDL-C, TDL-D, TDL-E \\ \hline
        Fading Distribution & Rayleigh\\ \hline
        \begin{tabular}[c]{@{}l@{}}Number of Antennas\\ (Tx, Rx)\end{tabular} & 1, 1\\ \hline
        Carrier Configuration & 51 RBs, 30 kHz SCS, Normal CP\\ \hline
        Sub-carriers Per RB & 12 \\ \hline
        Symbols Per Slot & 14\\ \hline
        Slots Per Subframe/ Frame & 2, 20\\ \hline
        Frame Duration & 10 ms\\ \hline
        NFFT & 1024\\ \hline
        Transmission Direction & Downlink \\ \hline
        PDSCH Configuration & PRB: 0-50, All symbols, Type A, 1 Layer \\ \hline
        Modulation & 16QAM \\ \hline
        DM-RS Configuration & \begin{tabular}[c]{@{}l@{}}Ports: 0, Type A Position: 2\\ Length: 1, Config: 2\end{tabular} \\ \hline
        Delay Spread & 1 - 300 ns \\ \hline
        Doppler Shift & 5 - 250 Hz \\ \hline
        SNR & -5 - 20 dB, steps of 2 dB \\ \hline
        Sample Rate & 30720000 \\ \hline
        Noise & AWGN \\ \hline
        Interpolation & Linear \\ \hline
        Data set shape &  \begin{tabular}[c]{@{}l@{}}[11264, 612, 14, 2] = [samples, \\ SCS$\times$RBs, symbols per slot, \\ real \& imaginary components]\end{tabular} \\ \hline

    \end{tabular}
    \label{tab:my_label}
\end{table}

\subsection{Simulation Results}
In this subsection, we present the results of our experiments evaluating the performance of GraphNet, compared to state-of-the-art approaches. We compare GraphNet against ChannelNet (another DL-based method for CE), Practical CE, and LS methods. The experiments are conducted in fading channels with the TDL-A delay profile under various scenarios involving different values of delay spread and Doppler shift. These tests aim to demonstrate the effectiveness of GraphNet across a range of realistic channel conditions. 

\begin{figure}
    \centering
    \includegraphics[width=0.8\linewidth]{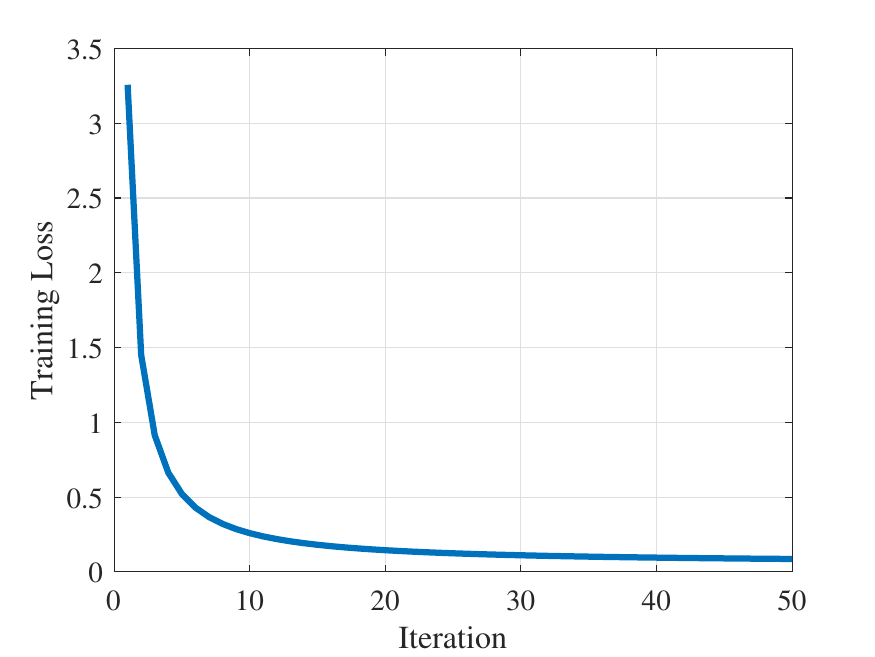}
    \caption{Learning performance of the proposed scheme}
    \label{MSESNR}
\end{figure}

Figure \ref{MSESNR} illustrates the total training loss of GraphNet, which is composed of the weighted loss of CE and noise estimation MSE. The loss starts at 3.5 and decreases to 0.25 within the first 10 iterations, stabilizing below 0.1 for the remaining iterations. This rapid decline and stabilization indicate the method's efficient convergence and strong performance.

\begin{figure*}[t!]
\centering
\subfloat[\footnotesize Delay Spread 3ns and Doppler Shift 5Hz ]{
\includegraphics[height=55mm]
{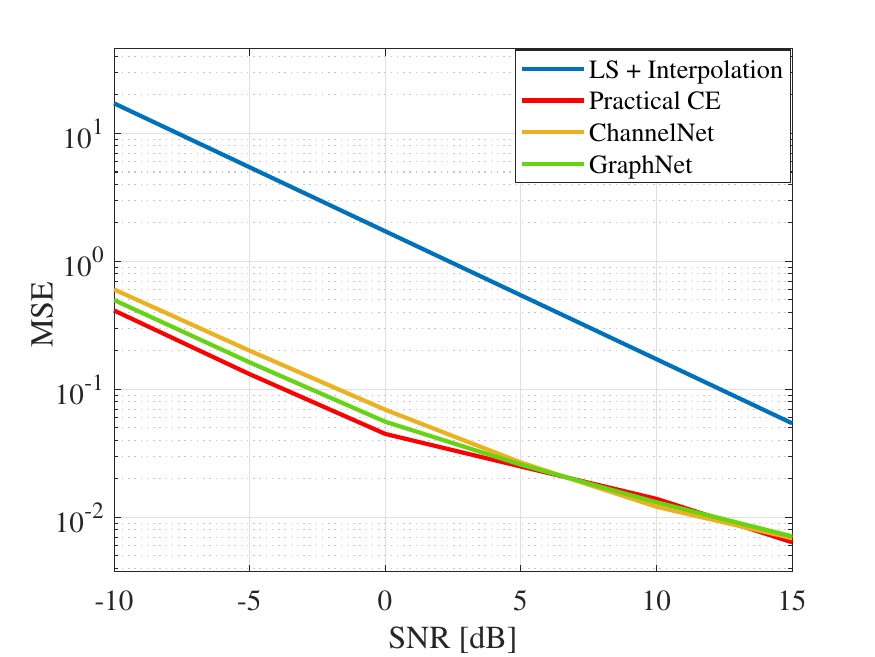}
\label{d3d5}
}
\hspace{32pt} 
\vspace{-.02in}
\subfloat[ \footnotesize Delay Spread 300ns and Doppler Shift 200Hz ]{
\includegraphics[height=55mm]
{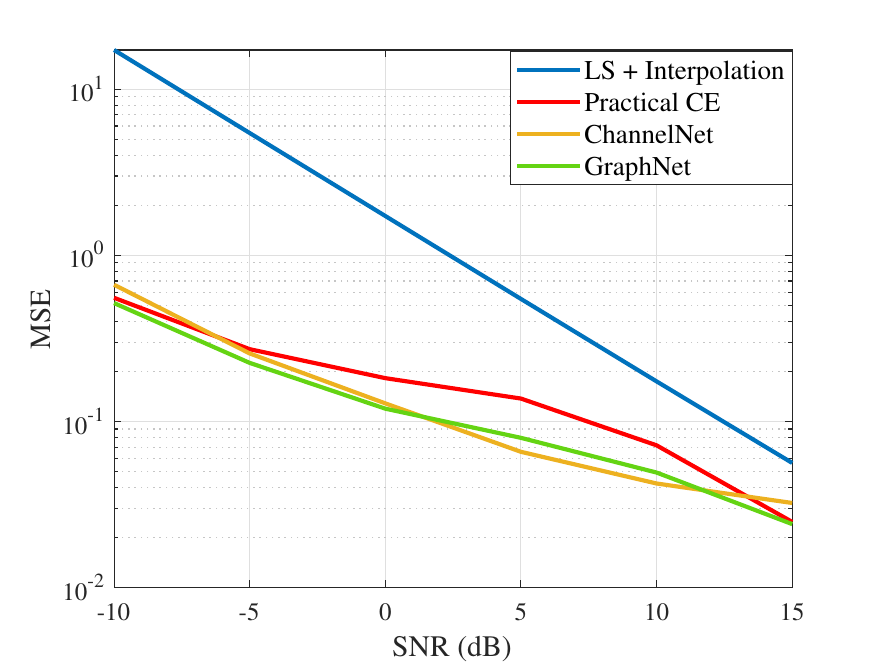}
\label{d300d200}
}
\caption{MSE performance versus SNR for Fading Channel with Delay Profile TDL-A }

\vspace{-.55cm}
\label{MSEvsSNR}
\end{figure*}

Figure \ref{MSEvsSNR} presents the MSE performance as a function of SNR, with SNR values ranging from -10 dB to 15 dB. First, we consider a scenario with a delay spread of 3 ns and a Doppler shift of 5 Hz (Fig. \ref{d3d5}), where we can see that all methods achieve similar performance, with Practical CE performing slightly better. However, in more dynamic environments where the channel varies rapidly—such as in Figure \ref{d300d200}, which considers a delay spread of 300 ns and a Doppler shift of 200 Hz—the performance of Practical CE significantly deteriorates.

This degradation occurs because the Practical CE method relies on assumptions of slow time variation and limited frequency selectivity, using techniques such as noise averaging and linear interpolation based on sparse \ac{dmrs} pilot positions. In high-mobility scenarios with large Doppler shifts and delay spreads, these assumptions break down: averaging fails to track rapid fluctuations, and interpolation becomes less accurate due to increased channel irregularity. Additionally, the LS-based initial estimation is highly sensitive to noise and lacks robustness against fast-fading effects. In contrast, ChannelNet and GraphNet exhibit strong performance, demonstrating DL-based methods' robustness to highly dynamic channel conditions by learning more flexible, non-linear mappings that generalize better to varying propagation environments.

\begin{figure*}[t!]
\centering
\subfloat[\footnotesize Delay Spread 3ns and Doppler Shift 5Hz ]{
\includegraphics[height=55mm]
{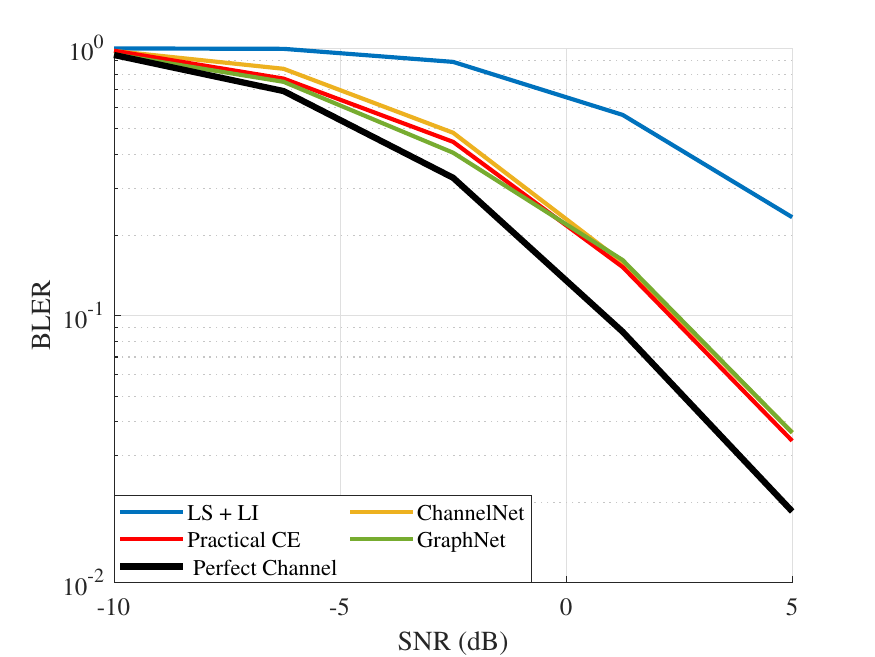}
\label{BLERd3d5}
}
\hspace{32pt} 
\vspace{-.02in}
\subfloat[ \footnotesize Delay Spread 300ns and Doppler Shift 200Hz ]{
\includegraphics[height=55mm]
{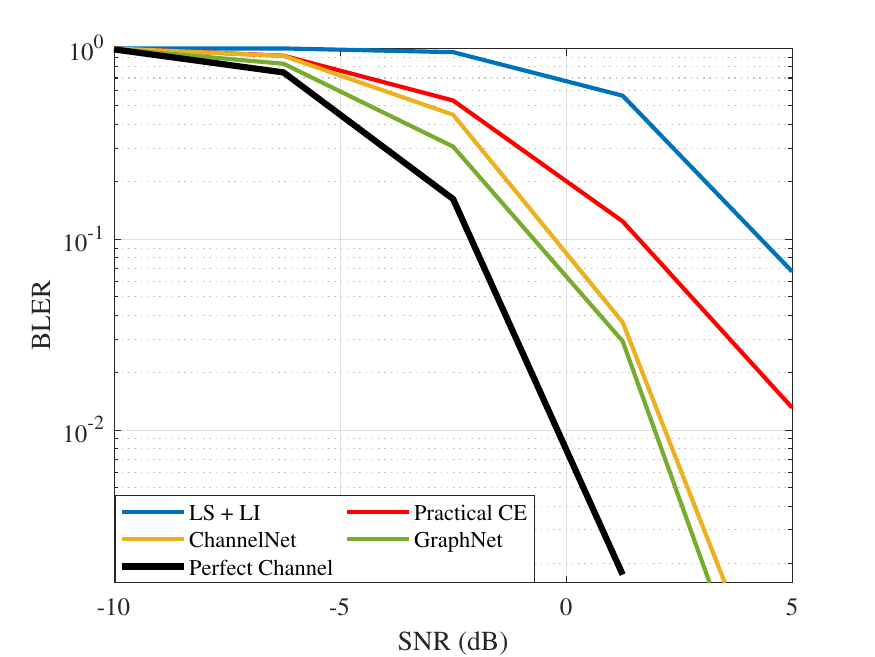}
\label{BLERd300d200}
}
\caption{BLER performance in Fading Channel with Delay Profile TDL-A }

\vspace{-.55cm}
\label{BLER}
\end{figure*}

Figure \ref{BLER} illustrates the \ac{bler} performance for the same channel conditions. All three evaluated methods perform similarly in the low delay spread and low Doppler shift scenario (Fig. \ref{BLERd3d5}). However, in the high delay spread and high Doppler shift scenario (Fig. \ref{BLERd300d200}), GraphNet outperforms both ChannelNet and Practical CE. While ChannelNet lags slightly behind, Practical CE experiences a severe performance drop due to its inability to track rapid channel variations effectively. One key reason for GraphNet's superior \ac{bler} performance is its incorporation of noise estimation. Although ChannelNet and GraphNet achieve similar MSE performance, and ChannelNet has a slightly lower average MSE, GraphNet delivers significantly better \ac{bler} performance. This suggests that precise noise estimation is crucial in improving \ac{bler} \cite{mendez2024bler}.

While Practical CE delivers strong performance in static or slowly varying channels, it is burdened by relatively high computational complexity, particularly due to the need for interpolation and averaging across both time and frequency domains. Although vectorized implementations have been proposed to improve efficiency \cite{9336351}, Practical CE still struggles to track rapid variations in highly dynamic environments with large delay spreads and high Doppler shifts. In such conditions, its performance degrades significantly. In contrast, DL-based estimators like ChannelNet and \ac{graphnet} offer significantly better robustness and adaptability. Notably, \ac{graphnet} not only achieves superior \ac{bler} performance—thanks to more precise noise estimation and graph-based feature extraction—but also exhibits a dramatically smaller model size and lower computational burden. As shown in Table \ref{simulation}, \ac{graphnet} contains only 2,307 parameters with a model size of 9.01 KB, compared to ChannelNet’s 574,244 parameters and 2.19 MB size. This makes \ac{graphnet} model size over 250 times lighter, making it an ideal candidate for resource-constrained or real-time 5G NR deployments, especially in rapidly varying channel conditions where Practical CE and heavier DL models fall short.

\begin{table}
\caption{SIMULATION PARAMETERS}
\label{simulation}
\centering
\begin{tabular}{|l|l|} \hline
\textbf{Parameter} & \textbf{Value} \\ \hline 
\multicolumn{2}{|c|}{\textbf{ChannelNet}} \\ \hline
CNN layers Num & 20 \\ \hline
\begin{tabular}[c]{@{}l@{}}Architecture \\ SRCNN\\ DNCNN\end{tabular}& \begin{tabular}[c]{@{}l@{}}
                                Input: (612, 14, 2) ,
                                Conv2D: 64 filters\\ 
                                Conv2D: 32 filters,
                                Conv2D: 2 filters \\
                                Conv2D: 64 filters + ReLU\\ 
                                Conv2D: 64 filters +BatchNorm + ReLU \\
                                (Repeated 15 more time),
                                Conv2D: 2 filters \\
                                \end{tabular} \\ \hline
Parameters & 574244 \\ \hline
Model size & 2.19 MB \\ \hline
\multicolumn{2}{|c|}{\textbf{\ac{graphnet}}} \\ \hline
Num nodes, edges $E$ & $8568$ , $25704$\\ \hline
Architecture   & \begin{tabular}[c]{@{}l@{}}
                                Input: (612, 14, 2),
                                SAGEConv: (2 $\rightarrow $ 32)\\ 
                                SAGEConv: (32 $\rightarrow $ 32),
                                SAGEConv: (32 $\rightarrow $ 2)
                                \end{tabular} \\ \hline
\begin{tabular}[c]{@{}l@{}} Architecture  \\  Noise estimation \end{tabular} & \begin{tabular}[c]{@{}l@{}} FC: linear (32 $\rightarrow $ 32), 
                            FC: linear (32 $\rightarrow $1) 
                                \end{tabular} \\ \hline
Parameters & 2307 \\ \hline
Model size & $ 9.01 $KB \\ \hline
\end{tabular}
\end{table}
\section{Conclusion}
In this paper, we propose GraphNet, a lightweight GNN-based estimator designed to enhance CE in 5G NR while also providing noise estimation. Through extensive evaluation across different channel conditions, including slow-varying and highly dynamic channels, GraphNet consistently outperformed traditional Practical CE and deep learning-based ChannelNet, particularly in fast-varying scenarios where Practical CE struggled. Beyond accuracy, GraphNet offers significant efficiency gains, achieving strong BLER and MSE performance with a model size over 250× smaller than ChannelNet. This makes it well-suited for real-world deployment in resource-constrained environments. In general, GraphNet is a robust and efficient solution for next-generation wireless systems, with promising potential for further adaptation and optimisation.

\section*{Acknowledgment}
The work presented in this paper was jointly funded by the Swiss National Science Foundation (grant number 219330) and the UK Department for Science, Innovation and Technology under the project YO-RAN.

\bibliographystyle{IEEEtran}
\bibliography{0_ref}

\end{document}